\documentclass[11pt]{article}
\usepackage{amssymb,cite,epsf,graphics,graphicx}
\begin{document}

\title{Conformal invariance and quantum integrability of sigma models on
symmetric superspaces.\\}
\author{\textbf{A.Babichenko}\thanks{%
e-mail: babichen@wicc.weizmann.ac.il} \\
\\
\textit{Racah Institute of Physics,}\\
\textit{\ the Hebrew University, Jerusalem 91904, Israel}.\\
 and \\
\textit{Department of Particle Physics,}\\
\textit{\ Weizmann Institute of Science, Rehovot 76100, Israel}\\}
\date{November 2006}
\maketitle

\begin{abstract}
We consider two dimensional non linear sigma models on few
symmetric superspaces, which are supergroup manifolds of coset
type. For those spaces where one loop beta function vanishes, two
loop beta function is calculated and is shown to be zero.
Vanishing of beta function in all orders of perturbation theory is
shown for the principal chiral models on group supermanifolds with
zero Killing form. Sigma models on symmetric (super) spaces on
supergroup manifold $G/H$ are known to be classically integrable.
We investigate a possibility to extend an argument of absence of
quantum anomalies in non local current conservation from non super
case to the case of supergroup manifolds which are asymptotically
free in one loop.
\end{abstract}

\section{Introduction}

Two dimensional (2d) non linear sigma models (NLSM) on
supermanifolds, with and without WZ term, seemed to be exotic
objects, when they appeared in condensed matter physics twenty
years ago as an elegant calculational tool in problems of self
avoiding walks \cite{PS} and disordered metals \cite{E}. Later on
they appeared in string theory context
\cite{S},\cite{Sh},\cite{AV}. A progress in their understanding
might be especially important for the theory of integer quantum
Hall plateau transition \cite{WZ},\cite{Z} and disordered systems
\cite{GLL} , but this progress is very slow. Many difficulties
prevent a usage of standard technique in investigation of 2d NLSM.
One of them is unavoidable non compactness of relevant target
space supermanifolds. Another -- a complicated representation
theory of the supergroups (their superalgebras), where so called
atypical representations play important, if not the main, role
\cite{GQS}. An interest to 2d NLSM on supermanifolds was renewed
recently in string theory in the context of ADS/CFT
correspondence, when it was understood that some ADS backgrounds
can be described in terms of supercosets \cite{MT},\cite{BVW}. For
example, $ADS_{5}\times S^{5}$ is nothing but ( bosonic part of)
the super coset $PSU(2,2|4)/SO(1,4)\times SO(5)$. Hyperactivity in
attempts to exploit integrability and methods of Bethe ansatz as a
calculational tool in checks of ADS/CFT correspondence (see e.g.
reviews \cite{BAR} and references therein), also supports this
interest, since both spin chains (on the gauge theory side) and 2d
NLSM (on the ADS side) appearing there, usually have a supergroup
symmetry. Some more examples of this kind appear in the context of
non critical strings ADS/CFT correspondence
\cite{P},\cite{V},\cite{CHZS}.

In this paper we try to investigate aspects of conformal
invariance and quantum integrability of 2d NLSM (without WZ terms)
on some symmetric supergroup manifolds. List of the models we are
interested in is the following. It starts from the principal
chiral models (PCM) on the basic supergroups Lie:
$G=A(m|n),B(m|n),D(m|n),D(2,1;\alpha ),G(3),F(4)$. In
addition we consider the following coset superspaces:%
\begin{eqnarray}
&&\frac{B(m|n)}{B(k|l)\times B(i|j)},\frac{B(m|n)}{D(m|n)},\frac{D(m|n)}{%
D(k|l)\times D(i|j)},\frac{D(m|n)}{A(m|n)},  \label{lis} \\
&&\frac{D(2,1;\alpha )}{A(1)\times A(1)\times A(1)},\frac{G(3)}{D(2,1;3)},%
\frac{F(4)}{C(3)}  \nonumber
\end{eqnarray}%
where $m=k+i,n=l+j$. In all these cosets the factor algebra $H$ is
a maximal regular subalgebra of $G$. Regular subalgebras of the
basic Lie
superalgebras were classified in \cite{VdJ}. For details of these
cosets embedding see Appendix. All the superspaces in (\ref%
{lis}) are symmetric. We hope that these toy models will serve as
a laboratory in investigation of more realistic ones, appearing
both in condensed matter physics, and in string theory.

One of the most interesting observations in this subject was done
in the paper \cite{BZV}, where it was shown that 2d NLSM without
WZ term (a PCM model) on the supermanifold with $PSL(n|n)$
symmetry is conformal in all orders of perturbation theory. In
\cite{BVW} this result was obtained for $PSU(2|2)$ PCM. The
authors pointed out the existence of a Casimir like chiral algebra
of the model, but a principal difficulties did not allow to
investigate the full spectrum of its representations. All the
machinery of CFT is hardly applicable for these non standard 2d
CFTs, although in some cases CFT methods were successfully applied
\cite{RS},\cite{SS},\cite{GQS}.

As it is well known, any 2d NLSM on a symmetric space is
classically integrable (see,e.g.\cite{AAR} and references
therein). Classical integrability expresses itself, in particular,
in the presence of conserved non local charges, or, in a more
rigorous way, in the presence of Backlund transform and spectral
parameter dependent Lax pairs. Generalization of the standard
procedure of non local current construction to the symmetric
superspace case seems straightforward. It was shown for ordinary
symmetric (non super) spaces that on the quantum level, absence of
anomaly in these non local current conservation is guaranteed only
if the factor group of a coset is either simple
\cite{AAG},\cite{L} or consists of a product of identical simple
group by itself \cite{EKY}. One can expect that the same feature
will remain in the case of symmetric superspaces. In this sense,
the list of cosets above represents a good candidates for quantum
integrable
models. (one should consider the first and the third cosets with $k=i$ and $%
l=j$).

Since the argument about presence/absence of anomaly in non local
currents is based on the dimensions of operators calculated as
engineering dimensions, one should be sure these dimensions are
correct in the UV limit. In the ordinary (non super-) case this is
guaranteed by asymptotical freedom (positiveness of the beta
function, at least in one loop) of 2d NLSM on symmetric spaces. As
we will see below, in general it is not the case for symmetric
superspaces. Requirement of asymptotic freedom which we are going
to impose in order to preserve an ability to talk about naively
calculated dimensions of the operators, will restrict possible
values of $m$ in the list above to be grater then $n$. So we start
from calculation of one loop beta functions for the above cosets.
As we will see, part of them
\begin{equation}
D(n+1|n),D(2,1;\alpha ),\frac{D(2n+1|2n)}{D(n+1|n)\times D(n|n)},\frac{%
D(n+1|n)}{A(n+1|n)},\frac{D(2,1;\alpha )}{A(1)\times A(1)\times
A(1)} \label{lisc}
\end{equation}%
have zero one loop beta function. We extended our calculations to
two loops and got zero. As we will show, the beautiful proof of
\cite{BZV} that beta function is zero in all orders of
perturbation theory, works in the same way for the superspaces on
the manifolds $D(n+1|n)$ and $D(2,1;\alpha )$. In \cite{BBHZZ} one
loop background field calculations of beta function were done for
some supercosets. Extension of all loops proof \cite{BZV} to the
supercosets seems problematical, but two loops beta function
calculation we made confirms that it is equal to zero. We
calculate the central charges of these cosets. Calculation of one
loop beta functions for the rest of the superspaces selects
asymptotically free ones. For them we analyze the quantum anomaly
in the first non trivial non local current conservation, and
conclude that there is no anomaly with a
proper choice of regularization. So the 2d NLSM on the superspaces (\ref{lis}%
) are quantum integrable, and moreover, those from the list
(\ref{lisc}) are conformal invariant.

\section{Beta function in one and two loops}

We start from a geometrical approach to background field
perturbation theory calculations of beta function for 2d NLSM on a
Riemannian supermanifold. We are going to discuss the action
\begin{equation}
S=\frac{1}{4\pi }\frac{1}{\lambda ^{2}}\int
d^{2}x\;Str[(G^{-1}\partial _{\mu }G)^{2}]  \label{ac}
\end{equation}%
where $G$ is an element of supergroup (supercoset) manifold, and
$Str$ is the supertrace. A review of the method and main results
for non super case one can find in \cite{K}. Recall that usual QFT
background field methods should be modified being applied to 2d
NLSM, if we wish to preserve target manifold Riemannian covariance
of calculations. One should expand the action around the classical
geodesic trajectory $\rho ^{a}$ on the manifold. Then a result of
calculations is expressed in terms of the basic covariant object
-- curvature tensor $R_{bcd}^{a}$, their covariant derivatives,
and products with different kind of indices contructions. In
particular, the one loop
beta function is proportional to the Ricci tensor%
\begin{equation}
\beta _{ab}^{(1)}=\frac{1}{2\pi }\lambda ^{2}R_{\;acb}^{c}=\frac{1}{2\pi }%
\lambda ^{2}R_{ab}  \label{b1}
\end{equation}%
In general, only the one loop result is regularization scheme
independent, higher loops depend on regularization. In dimensional
regularization there exists the choice, for which the two loop
result looks in the simplest way:
\begin{equation}
\beta _{ab}^{(2)}=-\frac{2}{3(2\pi )^{2}}\lambda
^{4}R_{a(cd)e}R_{\;\;\;\;\;\;b}^{e(cd)}  \label{b2}
\end{equation}%
where the parenthesis means the symmetrization over the indices,
and lowering/raising of indices is made by the manifold metric/its
inverse. In principle, all this technology of beta function
calculation may be extended to the supermanifolds. For definitions
of the main objects of Riemannian geometry on supermanifolds see
for example \cite{DW}. On the mathematical level of rigorosity,
there are some principal difficulties in basic definitions of
supermanifolds (even on the level of charts self consistency
\cite{Z1}). But there is a way to overcome these difficulties in
such a way
that usual objects of Riemannian geometry will be well defined \cite{B},\cite%
{R}. The only difference in these objects definitions from the non
super case is some extra minus signes related to the grade of
corresponding supermanifold coordinate. Carefully following all
the steps of covariant background field calculations described,
for instance, in \cite{K}, we got the same result for beta
function (\ref{b1}),(\ref{b2}), where $R$ now is the Riemannian
supercurvature tensor (particular expression in terms of structure
constants for supergroup manifold will be written below), and
additional minus signs, appearing in Feynmann diagrams as a result
of grade 1 fields loops propagators, are encoded in the
supermanifold metric used for indices contructions:
\begin{equation}
\widetilde{\beta }_{ab}=\frac{1}{2\pi }\lambda
^{2}R_{ab}-\frac{2}{3(2\pi )^{2}}\lambda
^{4}R_{a(cd]e}R_{\;\;\;\;\;\;b}^{e(cd]}  \label{bsup}
\end{equation}%
Here $(]$ means symmetrization if one of $c$ and $d$ is Grassmann
even, and it means anti symmetrization if both of them are
Grassmann odd. We hope to describe this technical calculational
details elsewhere.

We are going to apply this result to the supergroup manifold. The
basic supergroup structure is defined by its superalgebra
generators $Q_{A}$ with
structure constants $f_{\;BC}^{A}$:%
\[
\lbrack Q_{A},Q_{B}]=f_{\;AB}^{C}Q_{C}
\]%
where the commutator is graded $[A,B]=AB-(-1)^{\deg A\deg B}BA$.
In what
follows we denote the Grassmannian grade of the coordinate $\deg A$ as $|A|$%
. Since we are dealing with regular subalgebras $H$, the root
lattice of $H$
is a sublattice of the root lattice of $G$\footnote{%
Note that almost in all the supercosets from the list (\ref{lis})
one should use non distinguished Dynkin diagrams of $G$ for
realization of proper embedding of the factor subalgebra $H$ (see
Appendix).}. In other words, the whole set of generators of \ $G$
can be divided into two subsets $\{Q_{A}\}=\{Q_{a}\}\cup
\{Q_{i}\}$ -- generators of $H$ ($\{Q_{i}\}$), and generators of $G/H$ ($%
\{Q_{a}\}$). Following \cite{DW} (see also \cite{KY}, but in
another setting) one can derive the curvature on the coset
superspace in terms of the structure constants:
\begin{equation}
R_{\;bcd}^{a}=\frac{1}{2}f_{\;be}^{a}f_{\;cd}^{e}+\frac{1}{4}%
(-1)^{|b|(|c|+|d|)}f_{\;ce}^{a}f_{\;db}^{e}+\frac{1}{4}%
(-1)^{|d|(|b|+|c|)}f_{\;de}^{a}f_{\;bc}^{e}+f_{\;bi}^{a}f_{\;cd}^{i}
\label{cur}
\end{equation}%
Due to the main property of symmetric superspaces
$[Q_{a},Q_{b}]\subset H$,
all terms here but the last one, vanish. The Rici tensor defined as $%
R_{ab}=(-1)^{|c|(|b|+1)}R_{\;acb}^{c}$ is proportional to the
Killing form.
For the symmetric superspace case this property takes the form%
\begin{equation}
R_{ab}=-(-1)^{|c|(|a|+|b|+1)}f_{\;ai}^{c}f_{\;bc}^{i}=-(-1)^{|c|}f_{%
\;ai}^{c}f_{\;bc}^{i}=-K_{ab}  \label{ric}
\end{equation}%
An obvious but important for the future observation is that the
summation
over $i$ in the last formula may be extended on all the superalgebra $G$: $%
K_{ab}=(-1)^{|c|}f_{\;aD}^{c}f_{\;bc}^{D}$. On the other hand, it
is well known that the last expression, considered as a relation
on the whole superalgebra $G$, is nothing but $C_2\delta _{ab}$,
where $C_2$ is the value of the second Casimir operator evaluated
on the adjoint representation of the superalgebra $G$, i.e. the
dual Coxeter number of $G$: $R_{ab}\sim C_{2}\delta _{ab}$%
. The latter can be calculated purely algebraically, since it is
the value of the second Casimir operator on the adjoint
representation -- the rep. with highest weight with the highes
root of $G$. So we have an important simplifying statement:
\textit{the one loop beta function on symmetric (super)space
}$G/H$
\[
\beta _{ab}^{(1)}=-\frac{C_{2}}{2\pi }\lambda ^{2}\delta _{ab}
\]%
\textit{is proportional to the dual Coxeter number of
}$G$\textit{\ itself}. This statement is valid for non super
symmetric spaces as well. On the other hand, the value of the
second Casimir operator on any representation of quotient of a
(super)algebra $G$ by its regular subalgebra $H$ with highest
weight $\Lambda $, can be calculated using the formula
\begin{equation}
C=(\Lambda ,\Lambda +2\rho (G)-2\rho (H))  \label{cas}
\end{equation}%
where $2\rho (G)$ ($2\rho (H)$) is the sum of positive roots of
$G$ ($H$). Lets emphasis here that $2\rho (G)$ ($2\rho (H)$)
depend on the particular form of usually non distinguished Dynkin
diagrams one should use for the proper embedding of $H$ into $G$.
Change of the Dynkin diagram for a given superalgebra changes also
the order of its roots, and hence, the sum of positive roots.
Explicit case by case analysis confirms the statement we did
above: the contribution of the term $(\Lambda ,2\rho (H))$ with
$\Lambda $ the highest root of $G$, vanishes for all the cosets from the list (%
\ref{lis}). The values of dual Coxeter number for the basic Lie
superalgebras, which one can extract from, e.g. \cite{DIC}, we
list in the Table 1.$\qquad $

$%
\begin{tabular}{|l|}
\hline $G$ \\ \hline $C_{2}$ \\ \hline
\end{tabular}%
\begin{tabular}{|l|l|l|l|}
\hline $A(m|n)$ ($m\neq n$) & $A(m|m)$ & $B(m|n)$ & $C(m+1)$ \\
\hline $2(m-n)$ & $0$ & $2(m-n-\frac{1}{2})$ & $-2m$ \\ \hline
\end{tabular}%
$

$%
\begin{tabular}{|l|}
\hline $G$ \\ \hline $C_{2}$ \\ \hline
\end{tabular}%
\begin{tabular}{|l|l|l|l|l|}
\hline
$D(m|n)$ ($m\neq n+1$) & $D(n+1|n)$ & $D(2,1;\alpha )$ & $G(3)$ & $F(4)$ \\
\hline $2(m-n-1)$ & $0$ & $0$ & $6$ & $2$ \\ \hline
\end{tabular}%
$

Table 1.

From this table one can see that if we are interested only in
asymptotically free 2d NLSM on supergroups and their maximal
regular
supercosets (\ref{lis}) with a non positive beta function (non negative $%
C_{2}$), one should reject the supergroup $C(m+1)$. $A(m|n)$ ($m\neq n$) and $%
B(m|n)$ can be taken only with $m>n$, and $D(m|n)$ ($m\neq n+1$)
is acceptable if $m>n+1$. In addition
$A(m|m),D(n+1|n),D(2,1;\alpha )$ and their cosets from (\ref{lis})
are candidates for conformal field theories. Moreover the case
$A(m|m)$, which is the most popular in string oriented literature,
was proven to be really conformal field theory in all loops of
perturbation theory. We are going to concentrate on other cases.
As we said, the values of $C_{2}$ listed above are at the same
time the values for one loop beta functions for the cosets of
these supergroups (\ref{lis}). The
cases of PCM models on $D(n+1|n),D(2,1;\alpha )$, and 2d NLSM on (\ref{lisc}%
) are good candidates to be CFTs exactly as it happened to $A(m|m)$ \cite%
{BZV}. In order to check this statement one can try to calculate
the two loop beta function, using (\ref{b2}), for these supergroup
manifolds and their cosets from list (\ref{lisc}).

With this goal, it is useful to work in matrix representations of
the superalgebras and their cosets (see Appendix). It is
convenient to chose the defining representation of $D(n+1|n)$,
since in addition to the minimal dimension, it
gives an invariant non degenerate bilinear form on the algebra by $%
g_{AB}=Str(E_{A}E_{B})$, where $E_{A}$ are the supermatrices of
the algebra
generators. They can be chosen as follows (see \cite{DIC}):%
\begin{equation}
E_{IJ}=G_{IK}e_{KJ}+(-1)^{(1+\deg I)(1+\deg J)}G_{JK}e_{KI}
\label{gend}
\end{equation}%
Here $I,J,K=1,...,4n+2$, $\left( e_{IJ}\right) _{KL}=\delta
_{IK}\delta _{JL} $, and the ortosymplectic form in the
supermatrix block form is
\begin{equation}
G=\left(
\begin{array}{cc}
\widetilde{G} & 0 \\
0 & \overline{G}%
\end{array}%
\right)  \label{g}
\end{equation}%
The Grassmann even $2n\times 2n$ matrix $\overline{G}$ will be chosen as%
\begin{equation}
\overline{G}=\left(
\begin{array}{cc}
0 & 1_{n} \\
-1_{n} & 0%
\end{array}%
\right)  \label{g1}
\end{equation}%
and, for the moment, the $(2n+2)\times (2n+2)$ matrix
$\widetilde{G}$ we fix as
\begin{equation}
\widetilde{G}=\left(
\begin{array}{cc}
0 & 1_{n+1} \\
1_{n+1} & 0%
\end{array}%
\right)  \label{g2}
\end{equation}

One can easily write down the (anti)commutation relations for the
generators (\ref{gend}) and read off from them the structure
constants. It is a straightforward (but not trivial) exercise
(using Mathematica program) to check that not only (\ref{ric}),
but also the second term in (\ref{bsup}) calculated from these
structure constants through (\ref{cur}), vanishes. For
raising/lowering the indices necessary in (\ref{bsup}), one should
use the
metric $g_{AB}=Str(E_{A}E_{B})$ and its inverse defined on the generators (%
\ref{gend}).

The same (actually much easier) calculation may be done for the $%
D(2,1;\alpha )$ Lie superalgebra. Here one can extract the
structure constants for the generators
$T_{i}^{(a)},i=1,2,3,a=1,2,3$ (Grassmann even), and $F_{\alpha
\beta \gamma },\alpha ,\beta ,\gamma =1,2$ (Grassmann odd),
from the following (anti)commutation relations (see e.g. \cite{C})%
\begin{eqnarray}
\left[ T_{i}^{(a)},T_{j}^{(b)}\right] &=&i\delta _{ab}\varepsilon
_{ijk}T_{k}^{(a)},  \label{genal} \\
\lbrack T_{i}^{(1)},F_{\alpha \beta \gamma }] &=&\frac{1}{2}\sigma
_{\mu
\alpha }^{i}F_{\mu \beta \gamma },  \nonumber \\
\lbrack T_{i}^{(2)},F_{\alpha \beta \gamma }] &=&\frac{1}{2}\sigma
_{\mu
\beta }^{i}F_{\alpha \mu \gamma },  \nonumber \\
\lbrack T_{i}^{(3)},F_{\alpha \beta \gamma }] &=&\frac{1}{2}\sigma
_{\mu
\gamma }^{i}F_{\alpha \beta \mu },  \nonumber \\
\{F_{\alpha \beta \gamma },F_{\mu \nu \rho }\} &=&C_{\beta \nu
}C_{\gamma \rho }(C\sigma ^{i})_{\alpha \mu }T_{i}^{(1)}+\alpha
C_{\gamma \rho
}C_{\alpha \mu }(C\sigma ^{i})_{\beta \nu }T_{i}^{(2)}-  \nonumber \\
&&-(1+\alpha )C_{\alpha \mu }C_{\beta \nu }(C\sigma ^{i})_{\gamma
\rho }T_{i}^{(3)}  \nonumber
\end{eqnarray}%
where $\varepsilon $ is totally antisymmetric tensor, $\sigma
^{i}$ are the Pauli matrices, and $C=i\sigma ^{2}$ is the $2\times
2$ "charge conjugation"
matrix. The invariant metric $g$ on the algebra can be chosen as \cite{C}%
\begin{eqnarray}
g(T_{i}^{(a)},F_{\alpha \beta \gamma }) &=&0,  \label{gd} \\
g(F_{\alpha \beta \gamma },F_{\mu \nu \rho }) &=&C_{\alpha \mu
}C_{\beta \nu
}C_{\gamma \rho },  \nonumber \\
g(T_{i}^{(a)},T_{j}^{(b)}) &=&-\frac{1}{2\kappa ^{a}}\delta
_{ij}\delta ^{ab} \nonumber
\end{eqnarray}%
with $\kappa ^{1}=1,\kappa ^{2}=\alpha ,\kappa ^{3}=-1-\alpha $.
Again, the calculation of the beta function in one and two loops
with these structure constants and supersymmetric bilinear non
degenerate form gives zero.

If we now want to do the same calculations for the coset
superalgebras from the list (\ref{lisc}), we have to find matrix
representation which will give possibility to divide all the
generators of $G$ into two parts -- those which are generators of
$H$, and the rest ones. The latter are the
generators of the coset. It can be easily done for all the coset cases (%
\ref{lisc}). The
situation is the most simple in the case of the coset $\frac{D(2,1;\alpha )}{%
A(1)\times A(1)\times A(1)}$. Here we factor out all the Grassmann
even part of the algebra. It means that the coset contains only
the generators $F$
from (\ref{genal}). Simple calculation of one and two loop beta function (%
\ref{bsup}) through the structure constants gives zero.

Block structure of matrix realization of other two cosets
embedding from the list (\ref{lisc}) are explained in Appendix.
Again the calculation of one and two loop beta function using the
Mathematica gives zero.

\section{Conformal invariance in all orders of perturbation theory.}

The proof of conformal invariance \cite{BZV} in all loops of
perturbation theory can be repeated for the non coset
supermanifolds from the list (\ref{lisc}). Here we recall this
proof. In the background field method of beta function calculation
one starts from the action%
\begin{equation}
S[G]=\frac{1}{4\pi }\frac{1}{\lambda ^{2}}\int
d^{2}x\;Str[(G^{-1}\partial _{\mu }G)^{2}]  \label{accos}
\end{equation}%
and any element is represented as $G(x)=\widetilde{g}(x)G_{0}(x)$
with a
classical background field $G_{0}(x)$ and quantum fluctuations $\widetilde{g}%
(x)$. Then the current

\[
J_{\mu }=G^{-1}\partial _{\mu }G=G_{0}^{-1}\partial _{\mu }G_{0}+\widetilde{g%
}^{-1}\partial _{\mu }\widetilde{g}=J_{\mu
}^{0}+\widetilde{g}^{-1}\partial _{\mu }\widetilde{g}=J_{\mu
}^{0}+\widetilde{J}_{\mu }
\]%
and the action, after passing to the Lie superalgebra fields \bigskip $%
\widetilde{g}=e^{\lambda g}$, separates into three pieces%
\begin{equation}
S=S[\widetilde{g}]+S[G_{0}]+\frac{1}{2\pi }\frac{1}{\lambda
^{2}}\int d^{2}x\;Str[\widetilde{J}_{\mu }(\partial _{\mu
}G_{0})G_{0}^{-1}] \label{s1}
\end{equation}%
When passing to the Lie superalgebra from supergroup $\widetilde{g}%
(x)=e^{\lambda g(x)}$, the fluctuation current $\widetilde{J}_{\mu
}=e^{-\lambda g}\partial _{\mu }e^{\lambda g}$ can be expanded in series in $%
\lambda $%
\[
\widetilde{J}_{\mu }=\lambda \partial _{\mu }g+\frac{\lambda
^{2}}{2}\left[
\partial _{\mu }g,g\right] +\frac{\lambda ^{3}}{3!}\left[ \left[ \partial
_{\mu }g,g\right] ,g\right] +...
\]
In the same way, Lagrangian for self interacting part
$S[\widetilde{g}]$
takes the form%
\[
L[e^{\lambda g}]=\frac{1}{4\pi }\left\{ Str(\partial _{\mu
}g\partial ^{\mu
}g)+\frac{2\lambda ^{2}}{4!}Str(\left[ \left[ \partial _{\mu }g,g\right] ,g%
\right] \partial ^{\mu }g)+...\right\}
\]

Insertion of these relations into the (\ref{s1}) gives the action
for $g$ with the kinetic term $Str(\partial _{\mu }g\partial ^{\mu
}g)$ and two types of interaction terms. The first type is powers
of $\partial _{\mu }g$
each coming with the external field $G_{0}$, and the second -- powers of $%
\partial _{\mu }g\partial ^{\mu }g$ coming without external background
field. Each interaction vertex is proportional to the structure
constat coming from (graded) commutators.

In calculation of beta function we have to renormalize all 1PI UV
divergent diagrams. (Of course there are IR divergent diagrams,
but suppose they were regularized by inclusion of small mass
terms.) Standard power counting arguments lead to the naive
divergence formula $D=2-V_{1}$, where $V_{1}$ is the number of
vertices with the single derivative $\partial _{\mu }g$. Since
each single derivative $\partial _{\mu }g$ is coming together with
an external line of the background field, the naive divergence
formula takes the form $D=2-E$, where $E$ is the number of
external lines. Therefore, we should analyze all 1PI diagrams with
one or two external lines. Diagrams without external background
lines will renormalize the action $S[e^{\lambda g}]$. It leads to
the vertex and wave function renormalization which will not
be important for us since we are interested in diagrams with no external $g$%
-lines. Moreover, the renormalization procedure can be chosen in a
such way that the group structure of renormalized vertices does
not change and they remain proportional to the structure constants
also after the renormalization. So we should consider 1PI diagrams
with one external background line (Fig.1.a) and two such external
lines (Fig.1.b).
\begin{figure}[tbp]
\begin{center}
\epsfysize=0.8 true in \hskip 10 true pt \epsfbox{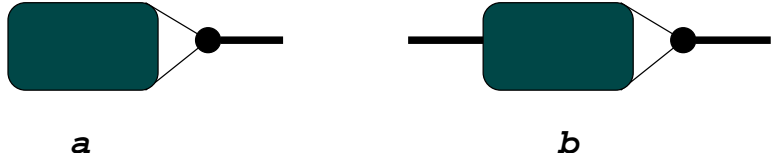} %
\end{center}
\caption{Structure of divergent Feynmann diagrams for background
field beta function calculation}
\end{figure}
The key observation is that the blob parts of both diagrams should
be supergroup invariant tensors, whereas the pulled off vertex
parts of diagrams are proportional to the structure constants.
Recall here that we call a tensor $t_{i_{1}i_{2}...i_{k}}$
supergroup $G$-invariant if
\begin{equation}
\sum_{s=1}^{k}\sum_{a\in G}\left( -1\right)
^{|a|}f_{\;bi_{s}}^{a}t_{i_{1}i_{2}...i_{s-1}ai_{s+1}...i_{k}}=0,\;\forall
b\in G,  \label{inv}
\end{equation}
In other words, the blob on the Fig.1.a is rank two supergroup
invariant tensor, and the blob on the Fig.1.b -- is rank three
invariant tensor. An obvious fact is that the \textit{only} rank
two invariant tensor is the non degenerate bilinear invariant form
of the Lie superalgebra. Recall that a natural way to chose this
form for the Lie superalgebras with the vanishing Killing form, as
the nominators in (\ref{lisc}), is $g_{ab}\sim Str(X_{a}X_{b})$,
where $X$ are the generators in the defining
representation for the $D(n+1|n)$ case, or the metric (\ref{gd}) for the $%
D(1,2;\alpha )$ case. In both cases it is a feature of these
superalgebras that the contraction of these metrics with the
structure constant, as it appears on the Fig.1.a, gives zero. In
order to calculate the contribution of diagrams Fig.1.b, one
should classify all the rank three superalgebra invariant tensors.
The full set of rank 3 tensors with indices $a,b,c$ can be chosen as $%
StrX_{a}StrX_{b}StrX_{c}$, $Str(X_{a}X_{b})StrX_{c}$, $%
Str(X_{a}X_{c})StrX_{b}$, $Str(X_{b}X_{c})StrX_{a}$, $%
Str(X_{[a},X_{b} X_{c]}) $, $Str(X_{(a},X_{b}X_{c)})$, where $()$
($[]$) means total (anti) symmetrization over indices $a,b,c$. Our
basis of generators is traceless, hence only last two tensors may
be non zero. Using the relation $GX=-X^{st}G$ vavid for
$D(n+1|n)$, one can easily show that the last tensor vanishes, as
well as all other odd rank totally symmetric invariant tensors.
For $D(2,1;\alpha)$ one can check this fact using explicit form of
the generators (\ref{genal}).
Hence we are left with only one invariant tensor of rank 3: $%
Str(X_{[a},X_{b}X_{c]})$ and it is proportional to the structure
constants. (Invariance of the structure constant as a tensor is
nothing but the Jacobi identity). It means that also all diagrams
of kind Fig.1.b vanish, since the contraction of two structure
constants they represent is proportional to the value of the
second Casimir in the adjoint representation (the Killing form),
which vanishes both for $OSP(2n+2|2n)$ and $D(2,1;\alpha )$. This
complete the proof of vanishing of beta function in all loops in
perturbation theory in $\lambda $.

Unfortunately we didn't succeed to generalize this proof of all
loops conformal invariance to the supercoset case.
But we consider one and two loops conformal invariance
discussed above as a good evidence for all loops conformal
invariance of the supercosets from the list (\ref{lisc}).

\section{Central charges}

A two dimensional quantum conformal field theory contains the
symmetry generator -- the energy momentum tensor
$T(z,\overline{z})$. The first question one asks, what is the
central charge of the CFT. We start from the cases of supergroup
manifolds. Here we follow the arguments of \cite{BZV}. We are
interested in exact non perturbative calculation of the
correlation function $TT$. First of all, as it was shown in
\cite{BZV} using the Zamolodchikov's equations for different
components of $T$, the coefficient $\alpha $ in the Sugawara
constructed component $T_{z\overline{z}}=\alpha
Str(J_{z}J_{\overline{z}})$, should be equal to zero, and hence
the energy-momentum tensor is
holomorphic. Then with a normalization $T_{zz}=-\frac{1}{2\lambda ^{2}}%
Str(J_{z}J_{z})$ we are interested in extraction of the central charge
from a non perturbative calculation of
\begin{equation}
\langle T(z)T(w)\rangle =\frac{1}{4\lambda ^{4}}\langle
Str(J_{z}(z)J_{z}(z))Str(J_{z}(w)J_{z}(w))\rangle
=\frac{c/2}{(z-w)^{4}} \label{c}
\end{equation}%
The crucial observation which enables a non perturbative
calculation is that in the expansion of $J_{z}$ all higher loops
come proportional to the contractions of the structure constants
with themselves, which give zero because of vanishing of the
Killing form (the dual Coxeter number) of the algebras
$OSP(2n+2|2n)$ and $D(2,1;\alpha )$. It means one can use just the
free fields $A$ instead of currents:
\begin{equation}
J_{z}\rightarrow \partial A,J_{\overline{z}}\rightarrow
\overline{\partial }A \label{ff}
\end{equation}%
with the action%
\begin{equation}
S=\frac{1}{4\pi \lambda ^{2}}\int d^{2}xStr(\partial _{\mu
}A\partial ^{\mu }A)  \label{ac1}
\end{equation}%
The propagator for the field $A$ should respect the supergroup
symmetry. For example, for the $OSP(2n+2|2n)$ case we can chose
\[
A_{IJ}=\left(
\begin{array}{cc}
a_{ij} & b_{i\overline{j}} \\
c_{\overline{i}j} & d_{\overline{i}\overline{j}}%
\end{array}%
\right)
\]%
with Grassmann even matrices $a,d$ and odd $b,c$, such that it
preserves the form
\[
G=\left(
\begin{array}{cc}
1_{2n+2} & 0 \\
0 & \overline{G}_{2n}%
\end{array}%
\right) ,\;\overline{G}=\left(
\begin{array}{cc}
0 & 1_{n} \\
-1_{n} & 0%
\end{array}%
\right) :\;A^{T}G=-GA,
\]%
which leads to
\[
a^{T}=-a,\;d^{T}=\overline{G}d\overline{G},\;b^{T}=-\overline{G}c,\;c^{T}=-b%
\overline{G}.
\]%
The following index structure of propagators respects this symmetry:%
\begin{eqnarray*}
\langle a_{ij}(z)a_{kl}(w)\rangle &=&\frac{1}{2}(\delta
_{il}\delta
_{jk}-\delta _{ik}\delta _{jl})(-\ln (z-w)) \\
\langle d_{\overline{i}\overline{j}}(z)d_{\overline{k}\overline{l}%
}(w)\rangle &=&-\frac{1}{2}\left( \delta _{\overline{i}\overline{l}}\delta _{%
\overline{j}\overline{k}}-\overline{G}_{\overline{i}\overline{k}}\overline{G}%
_{\overline{j}\overline{l}}\right) (-\ln (z-w)) \\
\langle c_{\overline{i}j}(z)c_{\overline{k}l}(w)\rangle
&=&\frac{1}{2}\delta
_{\overline{j}l}\overline{G}_{\overline{i}\overline{k}}(-\ln
(z-w))
\end{eqnarray*}%
Using these propagators, Wick theorem applied to (\ref{c}) with
substitution
of (\ref{ff}) gives the result%
\[
\langle T(z)T(w)\rangle =\frac{1/2}{(z-w)^{4}}
\]%
meaning that the central charge is $c=1$.

There is more compact way to do the same calculations. Using the
matrix realization basis $E_{IJ}^{m}$ for the superalgebra
$osp(2n+2|2n)$ described
above, one can write $A_{IJ}$ of the action (\ref{ac1}) as $%
A_{IJ}=\sum_{m}a_{m}E_{IJ}^{m}$. The propagator for $a_{m}$ is
just free field propagator $\langle a_{m}(z)a_{l}(w)\rangle
=-\delta _{ml}\ln (z-w)$.
We have to calculate%
\[
\langle Str(JJ)Str(JJ)\rangle =g_{ml}g_{pq}\langle (\partial
a_{m}\partial a_{l})(z)(\partial a_{p}\partial a_{q})(w)\rangle
\]%
where $g_{ml}=Str(E^{m}E^{l})$ is bilinear invariant and non
degenerate form on the superalgebra. Explicit calculation of the
previous expression reduces just to counting of diffenet
generators with the proper weights and gives the same answer. The
advantage of this method is obvious when we calculate the
supercoset central charge: one can use the last formula, but the
summation is running only over generators of $G$ which are not generators of $%
H$. This calculation for the PCM sigma model on
the $D(2,1;\alpha )$ supermanifold gives $c=1$.
For the cosets $\frac{D(n+1|n)}{A(n+1|n)}$ and $\frac{%
D(2n+1|2n)}{D(n+1|n)\times D(n|n)}$
we got $c=0$, and the coset The coset $\frac{%
D(2,1;\alpha )}{\left( A(1)\right) ^{3}}$ has $c=-8$.

Of course, the central charge says almost nothing about the two
dimensional CFT -- one has to know the full extended algebra of
the theory and its representations, i.e. the spectrum of the
primary fields. This problem doesn't seem solvable for today,
since almost nothing is known about CFTs where \textit{apriori
}there is no explicit factorization into holomorphic and
antiholomorphic parts, at least for the representations. As it was
shown in \cite{BZV}, one can construct holomorphic algebras, which
are believed
remain holomorphic anomaly free on the quantum level, i.e they are \textit{%
chiral} Casimir like algebras. The question is whether these
algebras contain all the symmetry of the theory. Another, more
realistic for solvability problem is an investigation of
representations of these chiral Casimir algebras at least on the
subset of holomorphic representations. Some steps in this
direction were done recently in \cite{GQS}.

\section{Quantum integrability}

Here we repeat the arguments \cite{L},\cite{AAG} about the
absence of quantum anomaly in the conservation of the first non
trivial non local current for UV asymptotically free sigma models
on the symmetric superspaces from the list (\ref{lis}) which are
not conformal, with either simple $H$, or semisimple consisting
from identical simple subalgebras. It means we are going to
reproduce the argument \cite{AAG} of quantum integrability of the
sigma models on the following supercosets:
\begin{equation}
\frac{B(2i|2j)}{B(i|j)\times B(i|j)},\frac{B(m|n)}{D(m|n)},\frac{D(2i|2j)}{%
D(i|j)\times D(i|j)},\frac{D(m|n)}{A(m|n)}\;(m\neq n+1),
\label{lis1}
\end{equation}%
\[
\frac{G(3)}{D(2,1;3)},\frac{F(4)}{C(3)}
\]%
with $i>j$ and $m>n$.

Recall that sigma models on symmetric spaces are always
classically integrable due to a possibility to construct a
parameter dependent flat connection. The same is true for
symmetric superspaces (see e.g. \cite{Klu},\cite{Y} and references
therein). In the same way as in the usual (not super) symmetric
spaces one can construct the first non
trivial non local conserved current \cite{AAR}:%
\begin{eqnarray}
Q^{(2)} &=&\frac{1}{2}\int dy_{1}dy_{2}\varepsilon
(y_{1}-y_{2})[j_{0}(t,y_{1}),j_{0}(t,y_{2})]-\int dyj_{1}(t,y)  \label{q} \\
\varepsilon (x) &=&\left\{
\begin{array}{c}
1,\;x>0 \\
-1,\;x<0%
\end{array}%
\right.  \nonumber
\end{eqnarray}%
The action of 2d NLSM on symmetric superspace $G/H$ is defined in
terms of
currents%
\begin{eqnarray*}
S &=&\frac{1}{4\pi ^{2}\lambda ^{2}}\int d^{2}xStr(k_{\mu }k^{\mu })=\frac{1%
}{4\pi ^{2}\lambda }\int d^{2}xStr(j_{\mu }j^{\mu }), \\
k_{\mu } &=&G^{-1}D_{\mu }G=G^{-1}(\partial _{\mu }G-GA_{\mu
}),\;j_{\mu }=-G^{-1}k_{\mu }G
\end{eqnarray*}%
where $A_{\mu }$ is the $H$ current. Equations of motion can be
written as
\[
D_{\mu }k^{\mu }=\partial _{\mu }j^{\mu }=0
\]

Potential source of anomaly on the quantum level is the first term in (\ref%
{q}), since product of operators at the same point requires \
regularization. One can say that there is no anomaly if there
exists a regularization procedure for $Q^{(2)}$ \ which preserves
its conservation. The arguments of \cite{L},\cite{AAG} are based
on the counting of all possible terms compatible with the
symmetries, which can appear in the operator product expansion of
two currents. The counting of such terms is possible if one can
trust to engineering dimensions of operators in the UV. This is
guaranteed in the asymptotically free theories, and that is why we
consider NLSM from list (\ref{lis1}) with the restriction
$i>j,m>n$. So the starting point is the splitting point
regularization of OPE
\[
\left[ j_{\mu }(t,x+\varepsilon ),j_{\nu }(t,x)\right]
=\sum_{k}C_{\mu \nu }^{(k)}(\varepsilon )Y^{(k)}(t,x)
\]%
where $Y^{(k)}$ is a complete set of local operators of dimension
not greater then two, such that $C_{\mu \nu }^{(k)}(\varepsilon )$
is divergent
or non zero when $\varepsilon \rightarrow 0$. All possible operators $%
Y^{(k)} $ should be consistent with the existing symmetries. The
left hand side is globally $G$-covariant and locally
$H$-invariant, so the same should be on the right hand side. One
can count all possible composite operators of dimension not
greater then 2 with these symmetry properties. If one chooses a
hermitian matrix realization for $G:$ $gg^{\dagger }=1$, any
operator of this kind can be written as $L_{1}gL_{2}g^{\dagger
}...L_{2k-1}gL_{2k}g^{\dagger }$, where $L_{i}$ is a product of
any number (including zero) of covariant derivatives $D_{\mu }$.
Not all of these operators are independent. There are no operators
of dimension 0 of this kind, there is one independent operator of
spin 1 -- it is $gD_{\mu }g^{\dagger }\equiv j_{\mu }$, and one
can chose two independent operators of dimension 2. The first is
$D_{\mu }D_{\nu }gg^{\dagger }\equiv gF_{\mu \nu }g^{\dagger }$,
and the second -- $D_{\mu }gD_{\nu }g^{\dagger }+gD_{\mu }D_{\nu
}g^{\dagger }\equiv \partial _{\mu }j_{\nu }$. Here $F_{\mu \nu }$
is the stress tensor in the subgrpoup $H$. The irreducible parts
of $gF_{\mu \nu }g^{\dagger }$ in the case of semisimple $H$, are
$G_{\mu \nu }^{(i)}$. We also used the fact that symmetric
superspace $G/H$ is by itself an irreducible representation of
$H$. The proof of this statement \cite{H} for the non super case
uses, in addition to the symmetricity property, the fact that $G$
is simple, and existence of invariant bilinear non degenerate form
on $G$. The only subtle point in copying of this proof on
symmetric superspace case is the last one: the natural form - the
Killing one - is identically zero for few cases of symmetric
superspaces. But as we said, there exists another bilinear form,
which is invariant and non degenerate. It can be successfully used
in this proof instead of the Killing form, and the proof still
works. Finally, the most general form of possible singular
terms in the Wilson expansion is%
\begin{equation}
\left[ j_{\mu }(t,x+\varepsilon ),j_{\nu }(t,x)\right] =C_{\mu \nu
}^{\rho }(\varepsilon )j_{\rho }(x)+D_{\mu \nu }^{\sigma \rho
}(\varepsilon )\partial _{\sigma }j_{\rho }(x)+\sum_{i}E_{\mu \nu
}^{(i)\sigma \rho }(\varepsilon )G_{\mu \nu }^{(i)}(x)  \label{w}
\end{equation}%
Moreover, because of identity $\sum_{i}G_{\mu \nu }^{(i)}=\partial
_{\mu }j_{\nu }-\partial _{\nu }j_{\mu }$ one can impose
\begin{equation}
\sum_{i}E_{\mu \nu }^{(i)\sigma \rho }(\varepsilon )=0  \label{e0}
\end{equation}%
Lorentz and PT invariance and charge conjugation of (\ref{w}) lead
to an ansatz for unknown coefficients $C,D,E$ with some properties
of their transformation under inversion of their argument sign. In
particular,
\begin{eqnarray*}
C_{\mu \nu }^{\rho }(\varepsilon ) &=&C_{1}(\varepsilon
^{2})g_{\mu \nu }\varepsilon ^{\rho }+C_{2}(\varepsilon
^{2})\left( \varepsilon _{\mu }\delta _{\nu }^{\rho }+\varepsilon
_{\nu }\delta _{\nu }^{\rho }\right) +C_{3}(\varepsilon
^{2})\varepsilon _{\mu }\varepsilon _{\nu }\varepsilon
^{\rho } \\
E_{\mu \nu }^{(i)\sigma \rho }(\varepsilon )
&=&E^{(i)}(\varepsilon ^{2})\varepsilon _{\mu \nu }\varepsilon
^{\sigma \rho }
\end{eqnarray*}%
Ward identity applied to (\ref{w}) gives first order differential
equations on the scalar functions $C_{i},D_{i},E^{(i)}$
\cite{AAG}. One can show that these equations have a solution. The
most singular part on the right hand side of (\ref{w}) is defined
by $C_{1}(\varepsilon ^{2})\sim 1/\varepsilon ^{2}$, and (\ref{w})
implies that the regularization
\begin{eqnarray*}
Q_{\delta }^{(2)} &=&\frac{1}{2}\int_{|y_{1}-y_{2}|>\delta
}dy_{1}dy_{2}\varepsilon
(y_{1}-y_{2})[j_{0}(t,y_{1}),j_{0}(t,y_{2})]-Z_{\delta }\int dyj_{1}(t,y) \\
Z_{\delta } &=&\int_{\delta }^{\Lambda }dxC_{1}(-x^{2})x
\end{eqnarray*}%
has a finite limit when $\delta \rightarrow 0$ ($\Lambda $ here is
an irrelevant IR cutoff). Explicit calculation of $dQ_{\delta
}^{(2)}/dt$ using (\ref{w}) and differential equations for
$C_{i},D_{i},E^{(i)}$, gives finally
\[
\frac{dQ_{\delta }^{(2)}}{dt}=\sum_{i}E^{(i)}\int dy\varepsilon
^{\mu \nu }G_{\mu \nu }^{(i)}(t,y)
\]%
One can see that in the case of simple $H$ we have zero on the
right hand
side just because of condition (\ref{e0}). Moreover, if we have semisimple $%
H $ composed as a tensor product power of one simple $H^{\prime }:$ $%
H=H^{\prime }\times H^{\prime }\times ...\times H^{\prime }$,
there is no reason why corresponding $E^{(i)}$ for each $H^{\prime
}$ should be different. Due to the same condition (\ref{e0}) we
again get zero. It completes the proof of absence of anomaly in
the non local current conservation for the symmetric superspaces
from the list (\ref{lis1}). Let us emphasis again that the anomaly
analysis presented here is a complete copy of the analysis for the
purely bosonic case, since the presence of the Grassmann odd
variables doesn't change any of the steps described above.

An important remark is in order here. As it was mentioned in
\cite{AAR},
if one computes the potential anomaly terms coefficients $%
C_{i},D_{i},E^{(i)} $ explicitly, they (including $C_{1}$) turn
out to be proportional to the dual Coxeter number of the group
$G$. Therefore the renormalization procedure scheme of $Q_{\delta
}^{(2)}$\ described above
doesn't work for supercosets with $G$ either $A(m|m)$, or $D(m+1|m)$, or $%
D(2,1;\alpha )$, since for them $C_{1}=0$. It shows that conformal
invariant cases are different and cannot be considered on the same
footing with other quantum integrable sigma models. At least role
of non local currents on the quantum level are different in the
conformal and integrable symmetric superspaces NLSM.

\section{Discussion}

We saw that beta function for symmetric superspaces of regular type (\ref%
{lis}) can be easily calculated in one loop. Those cases where the
beta function is negative (the value of the second Casimir
operator on the coset is positive), one can expect an asymptotical
freedom behavior. We have shown, that in the case of vanishing of
one loop beta function, both in the case of supergroup manifold
and in the case of supercoset spaces, the two loop beta function
vanishes. In the case of the supergroup manifolds, the proof of
\cite{BZV} of all loops vanishing of beta function on the
supermanifold $PSL(n|n)$, can be copied to the cases of supermanifolds $%
D(n+1|n),D(2,1;\alpha )$. It would be important to extend the all
loops conformal invariance proof to supercosets, and investigate
higher orders conformal invariance for other Ricci flat
supercosets, in particular, constructed recently in \cite{GM}.

One of the possible tools of analysis of these new two dimensional
CFT may be investigation of the chiral algebra of Casimir
operators. Recall, that by conjecture of \cite{BZV} the algebra of
operators
\[
W^{(k)}=t_{i_{1}i_{2}...i_{k}}J^{(i_{1})}(z)J^{(i_{2})}(z)...J^{(i_{k})}(z)
\]%
where $t$ is invariant tensor of rank $k$, remains chiral also on
the quantum level. It seems such a conjecture should be universal
for all the conformal NLSM (\ref{lis1}). Of course, there is a
question whether such algebras contain the full symmetry of these
models, but in any case, an understanding of their structure and
especially of their representations, including the spectrum of
primary fields, is an important and challenging problem. Another
interesting open question is the relation between these NLSM CFTs
and WZW models for the same supergroup (supercoset) manifolds
\cite{GQS}.

In the last section we saw that the standard arguments about the
absence of anomalies in non local current conservation known for
the symmetric non super cases may be extended to the super case.
It would be interesting to extend the all orders perturbation
theory analysis of non local currents conservation developed in
\cite{Be} to the NLSM considered here. Note that as we mentioned,
there is an obstacle on the way to treat conformal NLSM
(\ref{lis1}) on the same footing as integrable ones: the proof of
absence of anomaly in the non local current conservation doesn't
work, at least formally. Quantum integrability poses a question
about construction of their exact (relativistic) S-matrices. This
problem requires an understanding in which representation of $G$
leaves the fundamental massive multiplet of $G/H$ for all the
cases. One can also expect serious problems in the bootstrap
program realization, since, as usual, it goes in parallel to the
construction of a "fusion ring" of representations of algebra $G$,
and situation here is much more involved compared to the case of
non super spaces, mainly because of atypical representations,
which complicates the S-matrix construction (see e.g. \cite{BL}).
Of course the permanent problem of fixing of CDD ambiguity is also
there. All these interesting and not simple problems are important
in applications of 2d NLSM on supermanifolds to problems of
condensed matter and string theory.

\section{Acknowledgements}

I am thankful to D.Gepner, M.Gorelik and B.Noyvert for useful
discussions and to V.Kac, S.Ketov and C.Young for communications.
I am especially grateful to S.Elitzur for many productive
discussions and remarks. I am also thankful to Einstein Center for
financial support.

\section{Appendix}
\subsection{Embedding for cosets}
On the Fig.2 the supercosets embeddings are shown in terms of
superalgebras simple roots systems embeddings. Recall that we are
dealing with the regular subalgebras, when the root system of
subalgebra $H$ is embedded into the root system of $G$. For all
the supercosets from the list (\ref{lis}) the construction may be
described in terms of Dynkin diagrams as follows. One starts from
(usually not distinguished) extended Dynkin diagram of $G^{(1)}$.
The Dynkin diagram should be chosen in such a way that among
possible subdiagrams of $G^{(1)}$ one could find a diagram of $H$
by removing one (in the case of proper subalgebras) or more then
one (non proper case) nodes from diagram $G^{(1)}$. As we already
said, the full classification of such subalgebras (and hence
corresponding supercosets) was done in \cite{VdJ}. All cosets of
this kind as supergroup manifolds, turn out to be symmetric
superspaces. Using the data \cite{DIC} for full root systems of
Lie superalgebras, and the diagrams below for the simple roots,
one can list sets of positive and negative roots both for $G$ and
$H$. This data is necessary if one wishes to check the equality of
the value of second Casimir operator calculated on adjoint
representation of $G$ and on $G/H$ by the formula (\ref{cas}). The
positive roots system data is necessary also for much more tedious
exercise -- check that all the coset $G/H$ root systems are by
themselves irreducible representation moduli of $H$. As it was
mentioned above, this check is unnecessary, since the theorem of
this kind proved in \cite{H} can be extended on the super case.
\subsection{Matrix realization for cosets}
Here we explain the block structure of embedding for matrix
realization of cosets $\frac{D(n+1|n)}{A(n+1|n)}$ and
$\frac{D(2n+1|2n)}{D(n+1|n)\times D(n|n)}$. We start from the
former. We chose the standard matrix realization of $D(n+1|n)$ as
a $4n+2$ by $4n+2$ even supermatrix built from two even diagonal
blocks ($(2n+2)\times (2n+2)$ and $2n\times 2n$), and two non
diagonal Grassmann odd blocks. Generators were defined in
(\ref{gend}). We divide the matrix into 9 blocks. The rows and
columns are divided into 3 intervals: $I:[1,n+1]$,
$II:[n+2,3n+2]$, $III:[3n+3,4n+2]$. One can see that generators of
$D(n+1|n)$ which are non zero in the block $(II,II)$ (up to a
linear transform of diagonal Cartan subalgebra generators) form a
matrix realization of $A(n+1|n)$. Non zero entries of generators
in the blocks $(I,I)$, $(I,III)$, $(III,I)$, $(III,III)$ just
"copy" corresponding entries of the same generator located in the
block $(II,II)$. All of them should be reduced as elements of
factor algebra of the coset. The remaining generators of
$D(n+1|n)$ have non zero elements located in the blocks $(I,II)$,
$(II,I)$, $(II,III)$, $(III,II)$ and represent the generators of
the coset.

Similar but more involved "chessboard" embedding can be
constructed for the coset $\frac{D(2n+1|2n)}{D(n+1|n)\times
D(n|n)}$. Consider $n$ even case. We divide rows and columns of
the $(8n+2)\times (8n+2)$ even supermatrix realization of
$D(2n+1|2n)$ into 9 intervals: $I:[1,n/2]$, $II:[n/2+1,3n/2+1]$,
$III:[3n/2+2,5n/2+1]$, $IV:[5n/2+2,7n/2+2,]$, $V:[7n/2+3,9n/2+2]$,
$VI:[9n/2+3,11n/2+2]$, $VII:[11n/2+3,13n/2+2]$,
$VIII:[13n/2+3,15n/2+2]$, $IX:[15n/2+3,8n+2]$. One can see that
generators (\ref{gend}) with non zero elements in the blocks with
odd interval coordinate number -- $(I,I)$,
$(I,III)$,...$(III,I)$,... -- form the subalgebra $D(n+1|n)$,
whereas the generators with non zero elements in the blocks with
even interval coordinate numbers ($(II,II)$, $(II,IV)$,...) --
form subalgebra $D(n|n)$. Remaining generators with non zero
elements in the blocks with even-odd and odd-even interval
coordinates numbers represent the coset generators.

\newpage
\begin{figure}[tbp]
\epsfysize=6.0 true in \hskip 15 true pt \epsfbox{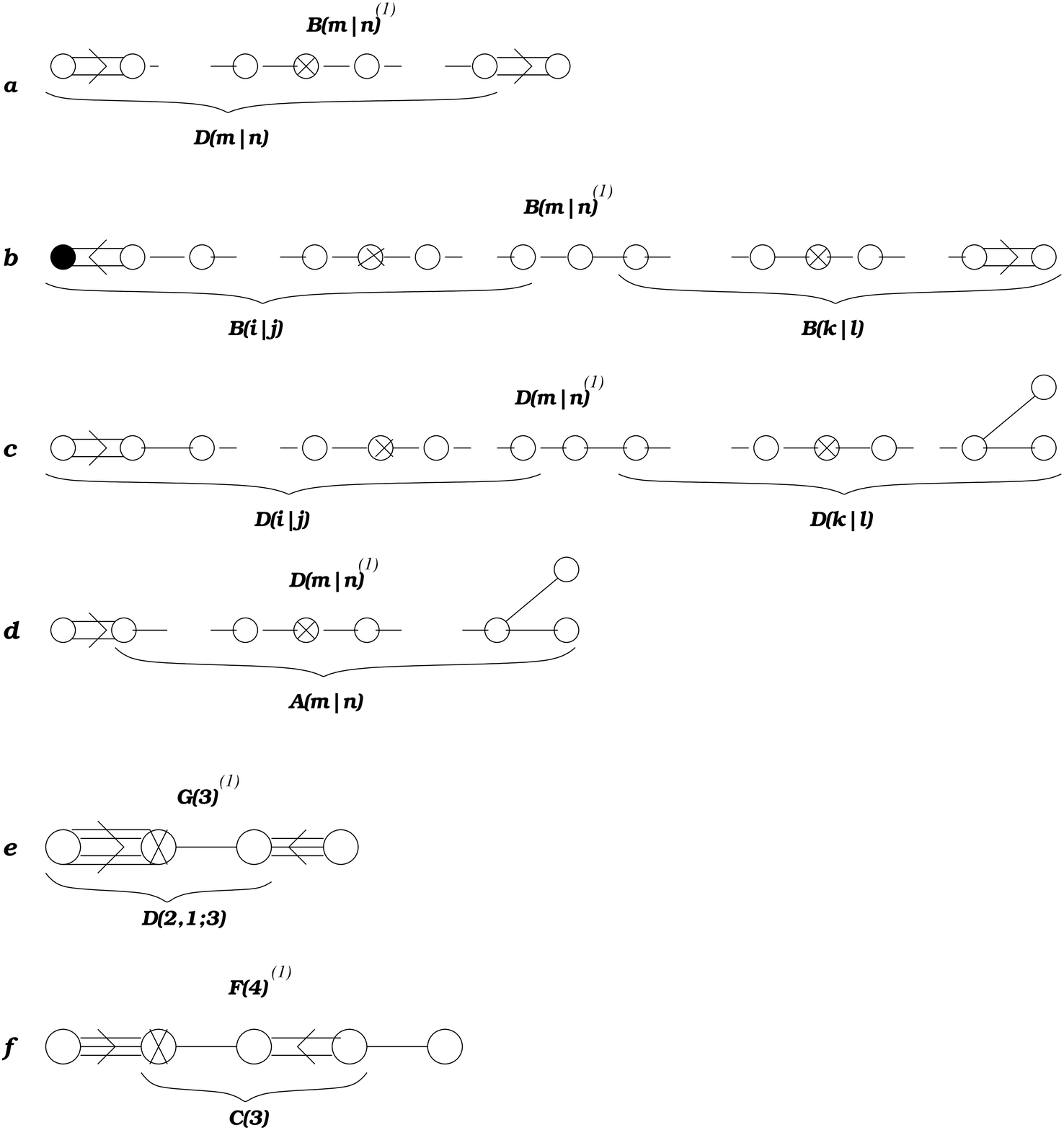}
\caption{Roots system embedding for supercosets. (a) --
$B(m|n)/D(m|n)$, (b) -- $B(m|n)/B(i|j)B(k|l)$, (c) --
$D(m|n)/D(i|j)D(k|l)$, (d) -- $D(m|n)/A(m|n)$, (e) --
$G(3)/D(2,1;3)$, (f) -- $F(4)/C(3)$.} 
\end{figure}

\end{document}